# Anchoring molecular rotors by on-surface synthesis

Kwan Ho Au Yeung, Tim Kühne, Frank Eisenhut, Francesca Moresco

Center for Advancing Electronics Dresden, TU Dresden, 01062 Dresden, Germany

**Abstract.** Single molecular rotor is an important component for constructing bot-tom-up molecular mechanical machines and a window for shedding light on complex physical and chemical questions about motions of organic molecules on surfaces. Stability of each component in such a molecular construction site is a crucial prerequisite. To realize a stable stepwise rotation of a molecule by a low temperature scanning tunneling microscope (LT-STM), atomic scale axles is particularly important. An ideal atomic scale axle is expected to balance between anchoring and mobility of rotating a single molecule on a metal surface under external excitations. In this Chapter, several chemical anchoring strategies on how to pin a molecular rotor are tested and discussed. Tip-induced manipulation and motion analysis are used as tools to investigate the properties and functionality of the proposed strategies.

## 1 Introduction

For building a molecular mechanical machine, the balance between anchoring and mobility is crucial for a single molecular rotor (schematically shown in Fig. 1). The first demonstration of a stepwise rotation of a molecular gear on an atomic scale axle was reported from Manzano and colleagues in 2009 [1]. This interesting strategy proposed to push a HB-NBP molecule onto an atomic defect by STM manipulation, which is natively presented on a herringbone elbow from the reconstruction of the Au(111) surface. Step-by-step and concentric rotations of individual molecules were presented, however, due to the long distance between such defects, no gears could be interlocked following this strategy. Another possibility was to produce a metal adatom by gently crashing the STM tip on the surface and to push the gear onto the adatom. Promising results were shown recently by Soe and colleagues, demonstrating rotations between a train of gears on Cu adatoms [2]. In short, the above strategies aim at constructing axles actively and manually after molecule deposition on sur-face. The possibility to anchor the molecules during deposition could be an alternative strategy.

Investigations on alternative strategies were attempted with the help of advanced synthetic chemistry. A rotating molecular rotor composed of a tripodal stator for vertical positioning, a star-shaped molecule for rotations, and a ruthenium atomic ball working as an axle was reported [3]. Taking this idea further from the view of the chemical bonding between the star-shaped molecule and ruthenium atomic ball, one can imagine that forming a new bond between an organic molecule and a metal sur-face could be a new direction by taking the advantage of on-surface synthesis.

Table 1 summarizes different anchoring strategies recently investigated by STM experiments that allows rotations of molecular gears on metal surfaces.

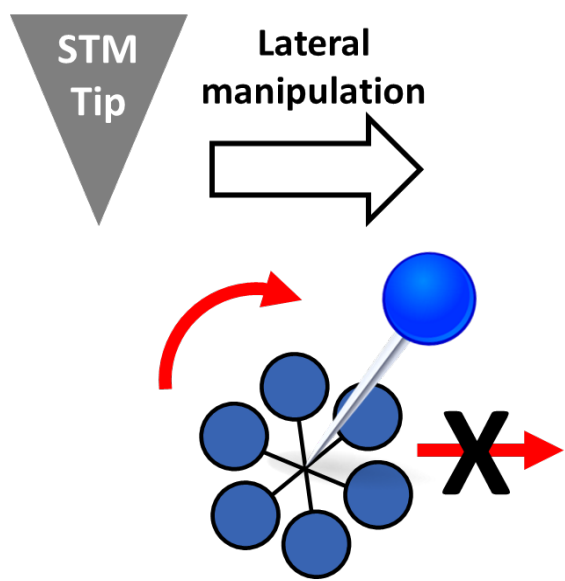

Fig. 1. Model representing an anchored/pinned molecular rotor rotated by a STM tip. An ideal anchoring strategy allows the rotational motion around the axle (blue pin) but hinders the lateral displacement.

In this Chapter, we discuss several chemical anchoring strategies on the Au(111) surface by employing a STM at low temperature (5 K) in ultra-high vacuum condition (UHV). STM allows not only the real-space visualization of molecules in atomic resolution, but also highly controlled manipulation by the tip apex. Importantly, tip-induced chemical reactions (i.e. dehydrogenation) are key factors for constructing anchoring sites by application of local voltage pulses in our studies. In the following sections, we present three different chemical anchoring strategies on several molecules with different sizes, symmetries and classes on the Au(111) surface: (Section 2) by a methoxy-group, (Section 3) by a methylamine-group, and (Section 4) by a de-hydrogenated dangling bond.

| Axle/ anchoring site | Description | Functionality of anchoring and corresponding motion of molecule | Reference |
|---|---|---|---|
| Atomic defect | By STM tip pushing the molecule on an atomic defect | Stepwise and stable rotation by lateral manipulation | Manzano et al [1] |
| Herringbone elbow | By STM tip pushing the molecule on a herringbone elbow caused by surface reconstruction i.e. Au(111) | Stepwise rotation and small lateral displacement by lateral manipulation | Manzano et al [1] |
| Metal adatom | By STM tip pushing the molecule on a metal adatom formed by tip dipping or deposition (i.e. Pd adatom on Cu surface) | Stepwise and stable rotation by lateral manipulation; Rotations between train of gears reported | Soe et al [2] |
| Molecular stator/ propeller | Deposition of firstly molecular stator on surface, then the deposition of molecular rotor with radical cyclopentadiene that leads to anchoring on the stator with a ruthenium atomic ball axle | Stable stepwise or continuous rotation depends on temperature; Chirality dependence for directionality induced by voltage pulse | Perera et al [3] Zhang et al [4] |
| Oxygen-/Sulfur-metal bond | Precursor molecules deposited on metal surface bound by oxygen or sulfur groups | Stable rotation excited by tip-induced voltage pulse; Chirality dependence for directionality induced by voltage pulse | Tierney et al [5] Eisenhut et al [6] |
| Dangling bond | By tip-induced or surface annealing to trigger dehydrogenation/dehalogenation of the molecules | Stepwise and stable rotation of dehydrogenated/dehalogenated molecule by lateral manipulation; Structure of molecule dependence; Up to three transmission of stable rotations | Au Yeung et al. (in preparation) |

Table 1. Overview of different reported anchoring strategies of molecular rotors/gears on metal surfaces performed by STM experiments. [4-6]

## 2  Anchoring molecular rotor by a methoxy group

Originally synthetized as an air-stable n-doping precursor molecule for the application in organic electronics [7, 8], we investigated the properties of single o-MeO-DMBI-I (2-(2-Methoxyphenyl)-1,3-dimethyl-1H-benzoimidazol-3-ium iodide) molecules deposited on the Au(111) surface. Our recent study shows that during thermal evaporation, the precursor is first reduced by iodine dissociation resulting in the radical form o-MeO-DMBI, and at temperatures around 490 K a C-O bond cleavage occurs, giving rise to DMBI-P and MeI as a byproduct. The resulted single DMBI-P molecules can be rotated stepwise and stable between six stable positions unidirectionally by application of local voltage pulses, depending on the molecule's chirality when adsorbed on Au(111) [9]. Hence, it is worth to investigate the surprisingly strong anchoring properties of the DMBI-P molecules and verify with motion analysis (Section 2.2). In this section, the quality of the anchoring strategy with the DMBI-P molecules (o-MeO-DMBI-I as precursors) will be discussed.

### 2.1  On-surface synthesis

We deposited the o-MeO-DMBI-I precursors at submonolayer coverage on a clean Au(111) surface kept at room temperature. After the C-O bond cleavage, the radical DMBI-P molecule chemisorbed on the Au(111) is anchored via back-donation (Fig. 2).

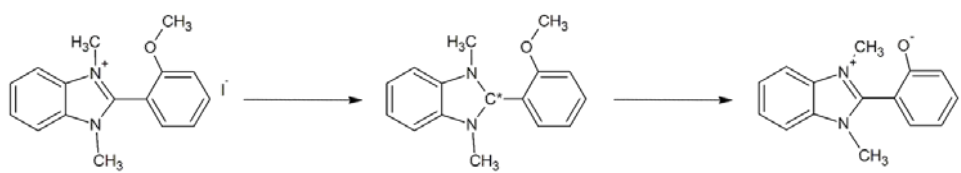

Fig. 2. Schematic diagram for expected products after on-surface synthesis by annealing. Iodine is first cleaved and then C-O bond cleavage occurs, forming a radical DMBI-P molecule.

### 2.2  Anchoring and rotation

The presence of a methoxy-group and of the radical form suggests a strong molecule-surface interaction that can be tested by pushing the single molecules with lateral manipulation. Upon manipulations, the molecules cannot be translated and no lateral displacements are observed. Manipulation trials with reduced tip height and con-sequently stronger tip-molecule interaction never lead to any lateral displacements. On the other hand, the orientation is clearly changed following the manipulation trajectory. This finding indicates that the rotation occurs around a fixed anchoring point at the side of the methoxy-group. In addition, the manipulation experiments demon-strate that the molecules are not physisorbed but chemically anchored to the Au(111) surface, as expected from the similar case of thioether-molecules which are known to strongly bind to surface atoms on coinage metals [10]. We therefore conclude that the single DMBI-P molecules are anchored to a gold atom through one of the two lone pairs of electrons of the oxygen (Fig. 3).

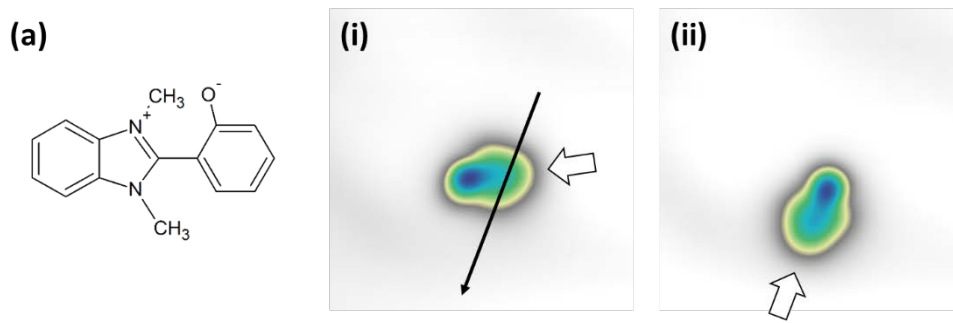

Fig. 3. The expected anchoring molecule DMBI-P adsorbed on the Au(111) surface. (a) Chemical structure of a DMBI-P molecule. (i)-(ii) Sequential STM images of rotating a DMBI-P molecule on Au(111) before and after lateral manipulation. The black arrow indicates the manipulation trajectory. (Parameters: I = 4.0 nA; V = 10 mV) (All image sizes: 5 nm × 5 nm; V = 0.2 V; I = 50 pA).

## 3    Tetracene-based rotor with a methylamine pin

The studies of poly-acenes on surface have been driving large attentions not only because of their instability starting from seven or more phenyl rings in solution, but more importantly to gain a quantum mechanical understanding and insights into their interesting electronic properties as extended π-electron structures in the one-dimensional limit. Recent advance regarding the on-surface synthesis of poly-acenes successfully reached to dodecacene consisting of twelve fused phenyl rings [11]. Other than the electronic properties, it is intriguing to understand the other physical and chemical properties, for instances, the adsorption on surface and the mechanical properties in terms of molecular mechanical machines with its extendable length in one dimension. More interestingly, large acenes show a very mobile characteristic, sometimes even a challenge for STM imaging from our observations (*i.e.* dodecacene). One could expect an anchoring or adsorption strategy for studying larger acenes in the future. The on-surface formation of tetracene and the corresponding electronic properties were reported in our earlier reports [12] [13]. In this section, we present a tetracene-based (Tn) rotor with a methylamine pin.

### 3.1    On-surface synthesis

After evaporating the tetracene precursors with two methylamine groups (Fig. 4), STM images reveal the topography of the individual molecules as a dumbbell shape. The protruding parts can be assigned to the methylamine groups of the molecule. Interestingly, after annealing at 443 K for five minutes on the Au(111) surface, four different species are observed. According to Fig. 4, we assign Cases 1 to 4 to the species for convenience. To confirm the atomic structures and the conformations of the resulted molecules, CO-functionalized tips were used to enhance the resolution of the STM images taken in constant height mode. For Cases 1 and 2, the STM images show two additional side groups on the same or opposite sides and for Case 3, only one side group is visible. Our findings suggest that the methylamine group can be either cleaved from one carbon atom leading to the formation of a benzene ring with a methylamine group at the side, or can be detached completely from the molecule leading to the formation of an acene. Note that the side groups can also be cleaved by voltage pulses above 3.0 V [12][13].

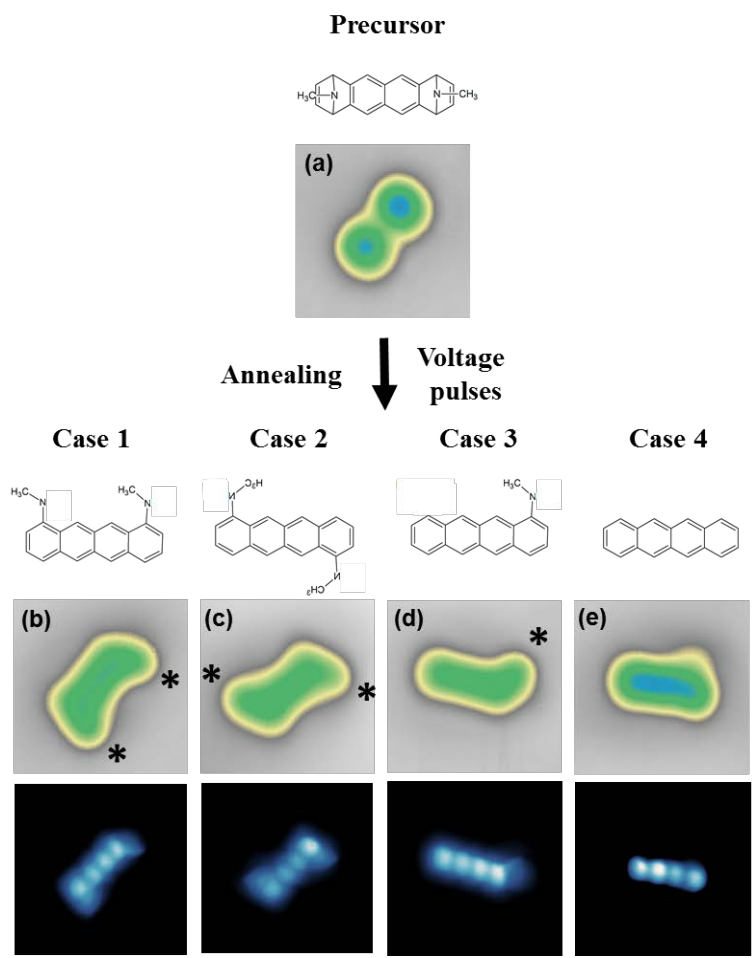

Fig. 4. The tetracene-based Tn(NCH$_3$)$_2$ molecular rotor. Top row: The proposed chemical structures of the precursor molecule and the resulting molecules after surface annealing at 170 °C or voltage pulses. (a) STM image of the precursor molecules on Au(111). (Image size: 2.5 nm x 2.5 nm; V = -560 mV; I = 73 pA) (b)-(e) Middle row: Constant-current STM images of the different species after annealing. (Image sizes: 2.5 nm x 2.5 nm; V = 500 mV; I = 100 pA); Bottom row: Corresponding constant-height STM images of the molecules achieved with a CO-functionalized tip. (Image sizes: 2.5 nm x 2.5 nm; V = 10 mV)

## 3.2 Rotational motion

As one can expect that a single anchoring site can work as an axle allowing for a stable rotation, we accordingly used the lateral manipulation mode of the STM to verify this. Indeed, it is not possible to move or rotate the molecules in Cases 1 and 2, while the ones in Case 3 can be rotated. Fig. 5 shows a sequential full rotation in six steps by lateral manipulation of the molecules from Case 3. No lateral displacements are observed during manipulations. The results indicate that a strong molecule-surface interaction was generated due to the nitrogen of the side group as an anchor. On the contrary, the Tn molecule without any side groups (Case 4) can be translated easily with a relatively low tunneling resistance of 5 MΩ. Our results reveal that it is possible to anchor a mobile tetracene molecule by a methylamine pin.

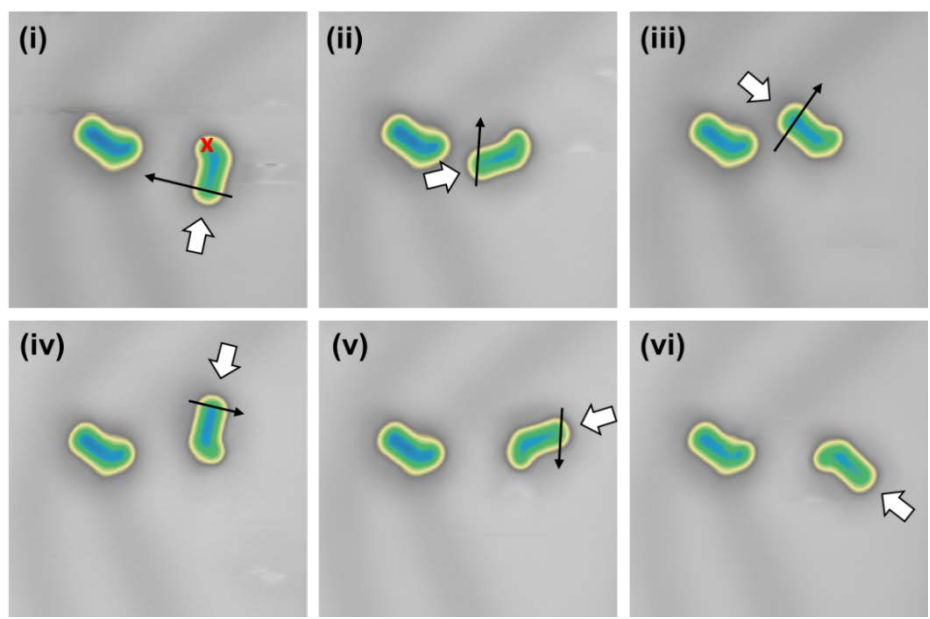

Fig. 5. Rotation of a reacted Tn(NCH$_3$)$_2$ molecule presented in Fig. 4 assigned as Case 3 on Au(111). (i)-(vi) Stepwise rotation of a tetracene-molecule with one methylamine-group by lateral manipulation. Manipulation trajectory is indicated by the black arrow. (All image sizes: 8 nm x 8 nm; V = 0.5 V; I = 100 pA; manipulation parameter: I = 2.0 nA; V = 10 mV)

## 4      Anchoring gear-like pentaphenylcyclopentadiene by dehydrogenation

In this section, we will explore the opportunity of dehydrogenated dangling bonds as an anchoring strategy within the framework of star-shaped pentaphenylcyclopenta-diene (PPCP; C$_5$Ph$_5$H) molecules on the Au(111) surface. As schematically shown in Fig. 6, the hydrogen atom (C-H bond) at the cyclopentadiene core is designed to work as a pin or an anchoring site of this molecule, taking a mechanical gear as an analogy. An ideal anchoring site should limit the freedom of lateral movement of the molecule. Hence, as a proof-of-concept, by inducing movements from the STM lateral manipulation, PPCP molecules have an advantage for us to analyze the pristine and radical states of the molecules on a metal surface. To implement the anchoring by dehydrogenated dangling bonds, the molecules are deposited by thermal evaporation on Au(111) and local STM tip-induced voltage pulses are used to trigger the dehydrogenation reaction for a concrete anchoring between the PPCP molecule and the Au(111) surface by a dangling bond.

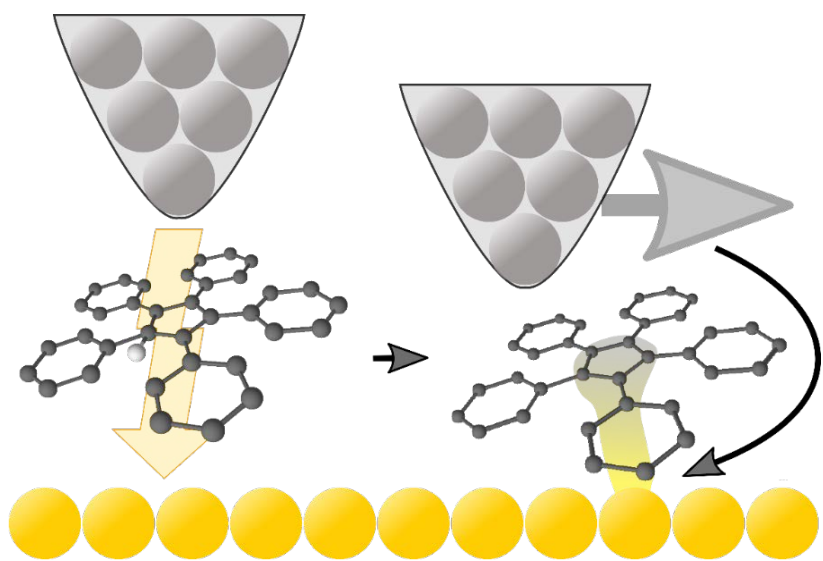

Fig. 6. Schematic diagram to illustrate the anchoring strategy for the gear-like PPCP molecule. The ball-stick model shows the chemical structure. After locally applying a STM tip-induced voltage pulse on a PPCP molecule at pristine state (left), a dehydrogenated dangling bond (a radical state) is formed with the Au(111) surface (right). To explore the properties of the anchor/pin, a tip-induced lateral manipulation is performed, usually at one of the phenyl rings of the molecule, to realize a rotational motion. STM images are taken before and after each manipulation for further motion analysis.

### 4.1   On-surface dehydrogenation

After evaporating PPCP molecules at 448 K for 30 s, the Au(111) surface was covered with a sub-monolayer film. Interestingly, clusters of PPCP molecules were found along the step edges or some small islands on the surface consisting of three or four molecules. This observation suggests that PPCP molecules are physisorbed on the Au(111) surface. In the STM images (Fig. 7), a clear feature of this molecule is the bulge (higher apparent height) near its center. If the five lobes that delineate the con-tour of the molecule are the five phenyl rings, this bulge should correspond to the hydrogen atom. Indeed, the application of a voltage pulse of about 3 V right above the bulge alters the conformation of the molecule. Repeated pulsing can flatten this bulge to a lower apparent height even if the first pulse does not change the conformation. Fig. 7 shows the STM image of two PPCP molecules where the upper one was flattened. (For simplicity, we call the flat/pulsed molecule as "radical" and the other as "pristine".). Peak-to-peak apparent height difference is about 0.7 Å. The flattening of the molecule, i.e. the removal of the central bulge, is a strong evidence for a successful dehydrogenation reaction. However, in some cases we observe different con-formations from the dehydrogenated molecule because of the many degrees of freedom of the phenyl rings.

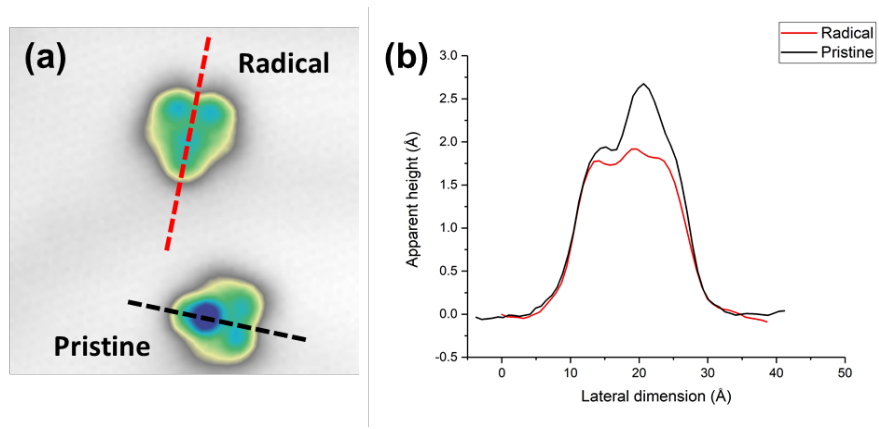

Fig. 7. (a) STM image and (b) corresponding height profiles of the pristine and radical PPCP molecules. Peak-to-peak apparent height difference is about 0.7 Å. (Image size: 6.3 nm × 6.3 nm; V = 500 mV; I = 20 pA)

### 4.2 Movement analysis

If a pristine PPCP molecule is dehydrogenated, it is reasonable to deduce that the remaining dangling bond can act as a pinning center so that the radical molecule rotates concentrically. Following this logic, a series of lateral manipulation experiments was performed. As shown in Fig. 8, it is possible to rotate the molecule step-by-step by lateral manipulation (Tunneling resistance = 25 MΩ). Note that some tiny lateral displacements occur after each manipulation, indicating that the anchoring is not strong enough for pure concentric rotations. To further understand the characteristics of the dehydrogenated dangling bond, we attempt to fix the molecule on one or two Au adatoms, however the bonding seems weaker than that on the flat gold surface.

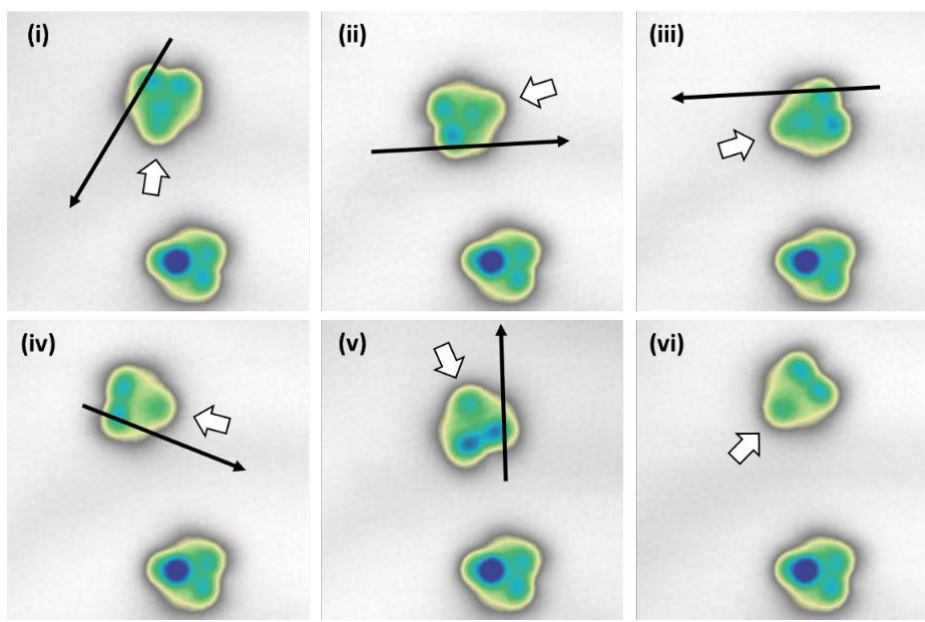

Fig. 8. Tip-induced movements of a radical PPCP molecule. (i)-(vi) The upper one is dehydrogenated by a 3.1 V voltage pulse while the lower one remains at the pristine state as a reference. The black arrows indicate the lateral tip manipulation events at V = 10 mV and I = 400 pA. (All image size: 6.3 nm × 6.3 nm; V = 500 mV; I = 20 pA)

### 4.3 Modified design from the framework of pentaphenylcyclopentadiene

A modification of the PPCP molecule with five further tert-butyl groups is shown in Fig. 9. Each additional tert-butyl group is connected to each phenyl ring (namely PPCP 2 in short). The molecule was designed to keep the single benzene rings in a fixed conformation, providing a more symmetric gear. When adsorbed on the Au(111) surface, the molecules can be found at the kink of the herringbone, usually individual ones instead of clusters. This suggests that the interactions between molecules are relatively weak. From the STM images, one can easily identify the five lobes around the center part which represent the tert-butyl-phenyl rings. The adsorbed PPCP-2 molecules can be rotated by STM lateral manipulation (Fig. 9 i-iii) without the need of voltage pulses, suggesting that the H atom can be already cleaved during evaporation, allowing also in this case a week anchoring of the molecule via the radicals 5-member ring. Note that it is still unclear to what extent a tert-butyl group can alter the tip manipulation (tip-molecule interactions) compared to a planar chemical structure (i.e. phenyl ring). In general, most of the cases suggest that a tert-butyl group is beneficial for lateral manipulation, which works similar to a "handle" [14] [15].

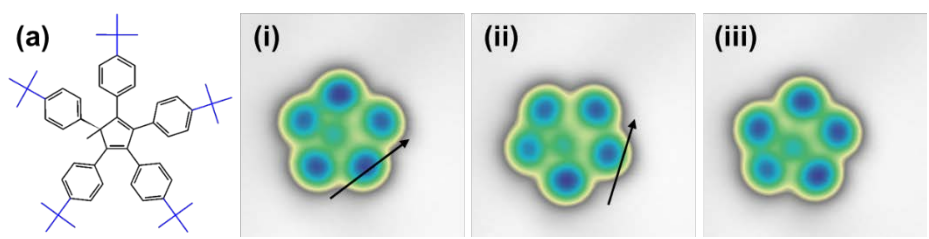

Fig. 9. Modified PPCP molecule (PPCP-2) with five tert-butyl groups attached to each phenyl ring. (a) Chemical structure of a PPCP-2 molecule. (i)-(iii) Sequence of rotations. The black arrows indicate the lateral tip manipulation events at V = 10 mV and I = 500 pA. (All image sizes: 5 nm × 5 nm; V = 500 mV; I = 5 pA)

Fig. 10 shows a train of PPCP-2 molecules that we formed to investigate collective transmission of motions. Four PPCP-2 molecules are brought together by tip manipulations from distinct sites on the Au(111) surface. After manually interlocking all the four molecules, it is not possible to translate any motions by manipulating on the side of the molecule (unlike the case from Fig. 9). Interestingly, driving the tip across the center lobe of the molecule can easily lead to lateral displacement. Such manipulation strategy is used for rotating three or more interlocked molecules. As shown in Fig 10, although it is not a stable and concentric rotation between molecules, a step of collective motion is clearly observed for the lower three molecules (green, yellow and red from the schematic drawing) transmitted from the uppermost molecule (blue). This indicates that the anchoring strategy using dehydrogenated dangling bonds in a radical state can provide both mobile and vertically stable motions, depending on the manipulation strategy.

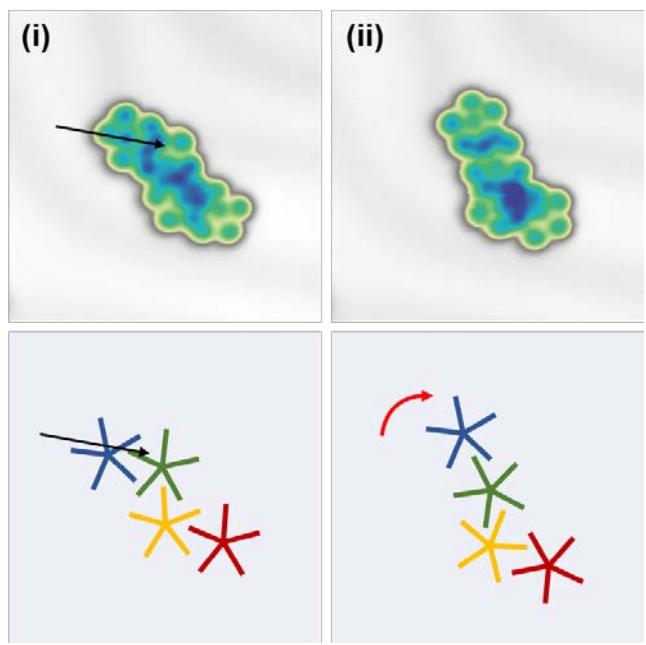

Fig. 10. Train of modified PPCP molecule (PPCP-2). (i)-(ii) Sequence of rotation between four molecular gears. The black arrows indicate the lateral tip manipulation events at V = 10 mV and I = 2200 pA. (All image size: 12 nm × 12 nm; V = 500 mV; I = 3 pA)

## 5    Summary and Perspectives

In this Chapter, we focus on the efforts on chemical anchoring strategies for different types of molecules by on-surface synthesis as an essential equipment of a molecular mechanical machine (*i.e.* a mechanical calculator). Three different classes of molecules, from one-dimensional tetracene and o-MeO-DMBI molecules to a two-dimensional gear-like PPCP molecule, are discussed. Among the proposed molecules and anchoring strategies, indeed, one cannot directly compare the quality of the anchors because of the differences between molecular structures (including the symmetries) that lead to different tip-molecule and molecule-surface interactions (even inter-molecular interactions for transmission of motion). Nonetheless, more importantly as a fundamental understanding of molecular anchors and rotors, it is crucial to explore the opportunity from in-solution chemical synthesis to on-surface synthesis and STM manipulation as a complex workflow. The promising strategies of anchoring are beneficial for future constructions of not only molecular mechanical machines, but also organic electronics and two-dimensional materials on surface in terms of the insights of the characteristics and functionalities of metal-organic complexes.

## Acknowledgments

This work has received funding from the European Union's Horizon 2020 research and innovation program under the project MEMO, grant agreement No 766864.

The authors thank Oumaima Aiboudi, Franziska Lissel, Claire Kammerer, Gwénaël Rapenne, and Diego Peña for providing part of the MeO-DMBI-I precursor molecules used in these studies.